\documentclass[11pt,preprint]{aastex}
\newcommand{\etal}{{et al.}}
\newcommand{\eg}{{\it e.g.,}}
\newcommand{\ie}{{\it i.e.,}}
\newcommand{\oiiilam}{[\ion{O}{3}] $\lambda5007$}
\newcommand{\oiii}{[\ion{O}{3}]}
\newcommand{\oiilam}{[\ion{O}{2}] $\lambda\lambda3726$,3729}
\newcommand{\oii}{[\ion{O}{2}]}
\newcommand{\ha}{{H$\alpha$}}
\newcommand{\hb}{{H$\beta$}}
\newcommand{\hg}{{H$\gamma$}}
\newcommand{\lya}{{Ly-$\alpha$}}

\received{2001 October 31}
\begin{document}

\title{The Extended Emission-Line Region of 4C\,37.43\footnotemark[1]}

\footnotetext[1]{Based on observations made with the NASA/ESA Hubble
Space Telescope, obtained at the Space Telescope Science Institute, which
is operated by the Association of Universities for Research in Astronomy,
Inc., under NASA contract NAS 5-26555.}

\author {Alan Stockton\altaffilmark{2,3,4}, John W. MacKenty\altaffilmark{5}, 
Esther M.\ Hu\altaffilmark{2}, and Tae-Sun Kim\altaffilmark{2,6}}

\altaffiltext{2}{Institute for Astronomy, University of Hawaii, 
2680 Woodlawn Drive, Honolulu, HI 96822}

\altaffiltext{3}{Research School of Astronomy and Astrophysics, Australian
National University, Canberra, ACT, Australia 2611}

\altaffiltext{4}{Visiting Astronomer, Canada-France-Hawaii Telescope, operated 
by the National Research Council of Canada, the Centre National de la Recherche 
Scientifique de France and the University of Hawaii.}

\altaffiltext{5}{Space Telescope Science Institute, 3700 San Martin Drive,
Baltimore, MD 21218}

\altaffiltext{6}{European Southern Observatory, Garching, Germany}

\begin{abstract}
We have explored the nature of the extended emission-line region around
the $z=0.37$ quasar 4C\,37.43, using extensive ground-based and {\it HST} 
imaging and spectroscopy.  
The velocity field of the ionized gas shows gradual gradients within
components but large jumps between components, with no obvious global
organization.  The {\it HST} \oiii\ image shows radial linear features
on the east side of the QSO that appear to mark the edges of an ionization
cone.  Concentrating on the bright emission peaks 
$\sim4\arcsec$ east of the quasar, we find through modeling that we require 
at least two density regimes contributing significantly to the observed
emission-line spectrum: one with a density of $\sim2$ cm$^{-3}$, having
essentially unity filling factor, and one with a density of $\sim500$ cm$^{-3}$,
having a very small ($\sim10^{-5}$) filling factor.  Because the temperatures
of these two components are similar, they cannot be in pressure equilibrium,
and there is no obvious source of confinement for the dense regions.  We
estimate that the dense regions will dissipate on timescales 
$\lesssim10^4$ years and
therefore need to be continuously regenerated, most likely by shocks.
Because we know that some QSOs, at least, begin their lives in conjunction
with merger-driven massive starbursts in their host galaxies,
an attractive interpretation is that the extended emission region comprises 
gas that has been expelled as a result of tidal forces during the merger
and is now being shocked by the galactic superwind from the starburst.
This picture is supported by the observed distribution of the ionized gas,
the presence of velocities ranging up to $\sim700$ km s$^{-1}$, and the
existence of at least two QSOs having similarly luminous and complex
extended emission regions that are known to have 
ultra-luminous IR galaxy hosts with current or recent starbursts.

\end{abstract}


\section{Introduction}

The first observation of emission lines from a resolved region around
a QSO was that of \citet{wam75}.  Other examples followed 
\citep{sto76,ric77,bor84,bor85,sto87}.
The origin of these extended emission-line regions (EELRs) around low-redshift
QSOs is still uncertain.  The distribution of the emission-line
gas generally appears to be uncorrelated with either the extended continuum 
emission or the radio structure, although both the incidence and the
luminosity of EELRs is higher for steep-spectrum radio quasars than for
flat-spectrum quasars or radio-quiet QSOs \citep{bor82,bor84,sto87}.  
Two principal suggestions have been
made:  (1) that the gas is debris from strong interactions or
mergers \citep{sto87}, and (2) that the gas is due to cooling flows from a 
hot surrounding medium \citep{fab87}.
We report here on an investigation of the EELR associated with
the 4C\,37.43.  This quasar has the most luminous EELR known
among QSOs at $z\lesssim0.5$ \citep{sto87}, making it an obvious candidate
for detailed study.  At the same time, the morphology and spectrum of
this EELR are similar to those of many other low-redshift QSOs, so the
results of this investigation should have a wider application.

\section{Observations and Data Reduction}
\subsection{Ground-Based Imaging and Spectroscopy}

We use several types of ground-based data on 4C\,37.43:  (1) imaging
in a 29 \AA\ bandpass centered on \oiiilam, to determine the morphology and
brightness distribution of the ionized gas; (2) imaging in the essentially
line-free continuum region between restframe 5050 \AA\ and 6050 \AA;
(3) imaging in a 290 \AA\
bandpass including \ha, along with an adjacent continuum bandpass, to
determine the \ha\ flux within an aperture equivalent to that used for
our {\it Hubble Space Telescope} ({\it HST}) FOS spectroscopy; 
(4) low-resolution spectroscopy to determine both the velocity structure of the
ionized gas and line ratios to compare with photoionization models
to estimate reddening; and (5) 
higher-resolution spectroscopy to measure the 
[\ion{O}{2}] $\lambda\lambda3726$,3729 doublet ratio
as a diagnostic of the electron density.

We obtained a total of 9000 s of narrowband [\ion{O}{3}] imaging with the
University of Hawaii (UH) 2.2-m Telescope and a Tektronix $1024\times1024$
CCD, through a filter centered at 6865 \AA\ and having a FWHM of 29 \AA.
In addition, we obtained 3000 s in an essentially line-free continuum bandpass
filter centered at 7606 \AA, with a FWHM of 1372 \AA.
The images were reduced via standard techniques, using flat-field exposures
of the illuminated interior of the dome, and the flux calibration was
determined from images of the spectrophotometric standard star Feige 67
\citep{mas88}.  Stellar objects have a FWHM of 0\farcs80 in the \oiii\
image and 0\farcs76 in the line-free continuum image.  For the brighter 
regions in both 
images, our S/N ratio is high enough to allow useful deconvolution, which we 
have carried out using the STSDAS contributed task ACOADD \citep[and references
therein]{hoo93}. 

We also obtained 10,800 s of \ha\ imaging at the UH 2.2-m 
Telescope, with a filter centered at 9131 \AA\ and having a 290 \AA\ FWHM 
bandpass.  Although redshifted \ha\ lies at $\sim9000$ \AA\, near the edge of
the filter bandpass, the transmission at this position was greater than 70\%.  
Line-free continuum exposures totaling 1500 s were obtained through a filter 
centered at 8338 \AA, with a FWHM of 880 \AA.  These FWHM for stellar objects
in these images was $\sim0\farcs9$.

Some of the low-resolution spectroscopy was obtained with image slicers 
\citep{sto00} and the
f/31 spectrograph on the UH 2.2-m Telescope.  The image slicers provided
3 contiguous slits on the sky and produced 3 independent spectra on the
Tektronix $1024\times1024$ CCD.  We tried two approaches:  half of the data
was obtained with a slicer having slit widths of 1\farcs15 at a position
angle (PA) of $-31\arcdeg$; the remaining half with a slicer having slit
widths of 2\farcs1 at a PA of $59\arcdeg$.  In both cases, the center slit
was centered on the bright emission peak about 4\arcsec\ east and slightly north
of the quasar. By using these image slicers, we
were able to collect essentially all of the flux from the main emission
region and eliminate any potential problem with atmospheric dispersion in
comparing the flux from \ha\ and \hb.  We obtained a total of 16,200 s of
useful integration.

We obtained additional low-resolution spectroscopy, with a total
integration of 3600 s, with the Low-Resolution
Imaging Spectrometer (LRIS; Oke \etal\ 1995) on the Keck II telescope.
With a 300 groove mm$^{-1}$ grating, the 1\arcsec-wide slit gave a FWHM of
$\sim12$ \AA.  The slit was placed through the QSO and the bright emission
region to the east at position angle 72\fdg7.

Our high-resolution spectroscopy of the [\ion{O}{2}] doublet was obtained
with the UH Faint-Object Spectrograph and a Texas Instruments $800\times800$
CCD on the Canada-France-Hawaii Telescope (CFHT).  Using a 1200 groove mm$^{-1}$
grating and a 1\farcs4 slit, we had a dispersion of 0.7 \AA\ pixel$^{-1}$
and an instrumental FWHM of 3.3 pixels, or 2.3 \AA.  
We obtained 7200 s total integration
centered on the peak of the emission-line region at slit PA $-34\arcdeg$.
We also obtained additional lower-resolution (FWHM 5.0 \AA) 
spectroscopy during the same
run, using a 500 groove mm$^{-1}$ grating and an image slicer with three
0\farcs7 slits on 1\farcs4 centers and at position angle 75\fdg5.  
By using 4 interleaved exposures
(3 of 2700 s duration, 1 of 2100 s), we were able to completely cover a 
region of $8\farcs4\times35\arcsec$.  The main purpose of these observations
was to determine the velocity field of the ionized gas over most of the
emitting region through measurements of the \oiiilam\ line.

\subsection{HST Spectroscopy and Imaging}

Our HST observations of 4C\,37.43 consist of Cycle 2 spectroscopy with the 
Faint-Object Spectrograph (FOS) and Cycle 6 imaging with the 
Wide-Field---Planetary Camera (WFPC2).  

The original objective of the FOS 
spectroscopy was to determine the \lya\ flux, for comparison with groundbased
observations of Balmer-line fluxes. The entrance
aperture was $4\farcs3\times4\farcs3$, but the height of the linear diode
array limited the effective aperture to $4\farcs3\times1\farcs4$, where the
4\farcs3 dimension was parallel to the dispersion axis.  For an aperture
of this size, the pre-refurbishment {\it HST} optical quality was
sufficiently good that the residual spherical aberration had no significant
impact on our results.  The center of this
effective aperture was offset from the quasar by 4\arcsec\ in position angle
$59\arcdeg$, to the peak of the extended emission.  In order to 
minimize the amount of scattered light from the wings of the
aberrated quasar PSF entering the spectrograph, we used an engineering mode
that allowed us to deflect the electron beam within the Digicon
detector so that only the lower third of the square entrance aperture (that
closest to the quasar) was projected onto the readout array.  
A total of 5550 s of exposure was obtained in two orbits.

The FOS observations were reduced with the improved STScI ``calfos''
pipeline in 1994.  This version included geomagnetic correction.  In
addition, a manual correction was applied for internal scattered light,
although this had only a minor impact on the resulting spectrum.  From
the absence of broad wings on \lya\ and \ion{C}{4} $\lambda1549$ and 
comparison with the line ratios observed in the FOS nuclear spectrum
\citep{bah93}, we established that any contribution of light from the 
quasar itself was small, in agreement with our estimate of the expected
light in the wings of the quasar PSF.

The WFPC2 imaging comprised both WFC3 and PC1 observations in the F814W 
filter, which is dominated by continuum radiation (although \ha\ falls
on the longward wing of the filter profile, where the transmission is
$\sim10$\% of the peak), and WFC2 observations with
a linear-ramp filter (LRF) centered on the redshifted 
[\ion{O}{3}]~$\lambda5007$ emission line.
The WFC3 F814W imaging consisted of two exposures at each of four dither
positions, with a total of 6400 s; the PC1 F814W imaging, of four exposures 
at each of four dither positions, with a total of 2560 s; and the WFC2 LRF 
imaging, of two exposures at each of two dither positions, with a total of
9200 s.  

After standard pipeline reduction, the images were combined using the
STSDAS {\it drizzle} procedure \citep{fru97}.  Cosmic rays were
first identified using the STSDAS {\it crrej} task, with careful adjustment of
the parameters to avoid any significant compromising of real data, and
masks for the cosmic rays were combined with the pipeline-generated data-quality
files to produce the masks used to reject pixels in the final combination.

\begin{figure}[p]
\epsscale{1.0}
\plotone{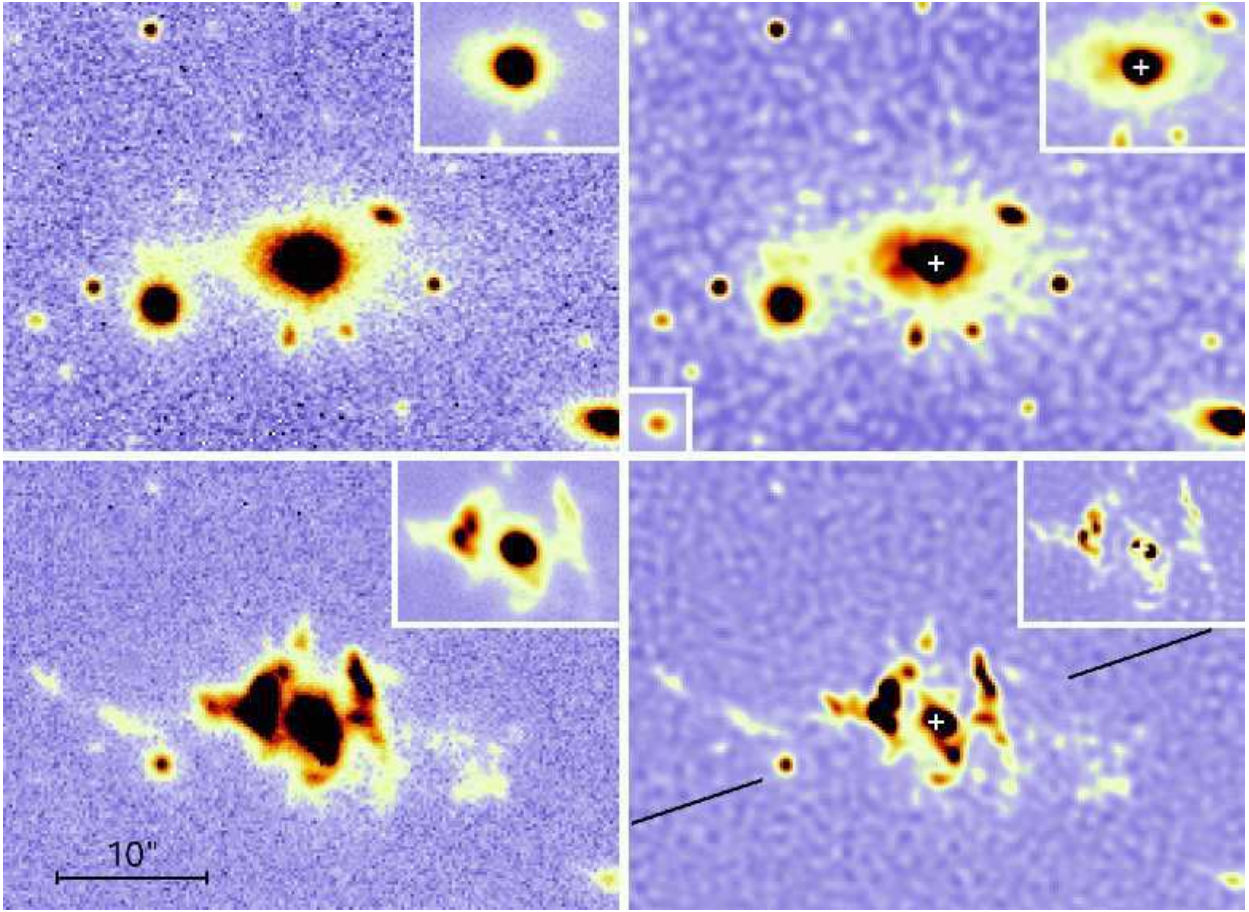}
\figcaption{Ground-based images of 4C\,37.43 obtained with the UH 2.2 m telescope.
The upper panels show an essentially line-free continuum image in a bandpass
covering rest-frame 5050 \AA\ to 6050 \AA.  The lower panels show an image
obtained in a 30 \AA\ bandpass filter centered on redshifted \oiiilam.  The
left panels show the original images, and the right panels show images
that have been deconvolved and have had the PSF removed with {\it plucy}
\citep{hoo94}.  The white crosses in the right panels show the positions
of the quasar.  The insets in the upper-right corners of each panel show
lower-contrast versions of the region around the quasar, and the small
inset in the lower-left corner of the upper-right panel shows the center of
the host galaxy at still lower contrast.  The lines in the lower-right panel
indicate the directions to the radio lobes.  The emission region referred to
in the text as ``E1'' consists of the double-peaked structure seen
$\sim4\arcsec$ to the left of the quasar in the lower panels.  Note also
the strong peak only $\sim0\farcs5$ to the right of the cross in the lower-right
panel.  North is up and East to the left
for this and all other images.\label{gbimages}}
\end{figure}

\section{Results}
\subsection{The Morphology of the Ionized Gas}
Figure~\ref{gbimages} shows our ground-based images in essentially line-free 
continuum and
in [\ion{O}{3}] emission, obtained with the UH 2.2 m telescope.  The most 
striking impression is how {\it unlike} the continuum and line images are:
the ionized gas seems to know virtually nothing of the distribution of the 
continuum
(presumably stellar) material.  The continuum extension to the east,
possibly forming a tidal bridge to the companion galaxy (which is known to
be at the same redshift as the quasar; \citealt{sto73,sto78}), has been 
described previously by \citet{sto86}, \citet{hic87}, \citet{sto87}, and
\citet{blo91}.  The strong emission-line peak $\sim4$\arcsec\ east-northeast 
of the quasar was first noted in $R$-band imaging (which
includes the strong [\ion{O}{3}] lines) by Stockton (1976); a lower-resolution
image in a narrow-band filter centered on [\ion{O}{3}] $\lambda5007$ has been
given by \citet{sto87}, and integral-field spectroscopy has been carried
out by \citet{dur94} and \citet{cra00}.  This emission structure, which we 
shall henceforth refer to as ``$E1$'', is the subject of much of the remainder 
of this paper.

Particularly in the deconvolved, PSF-subtracted \oiiilam\ image 
(lower-right panel of Fig.~\ref{gbimages}),
considerable new detail is seen in the structure of the ionized gas.  $E1$,
previously seen simply as
an elongated region, is now clearly double, with each of the peaks sporting
suggestive tail-like appendages.  The long linear structure to the northwest of
the quasar takes on an undulating appearance and breaks up into a series of
knots.  Finally, there appears to be a previously undetected emission peak 
even brighter than those to the east just 0\farcs5 west and slightly south 
of the quasar.

\begin{figure}[p]
\epsscale{1.00}
\plotone{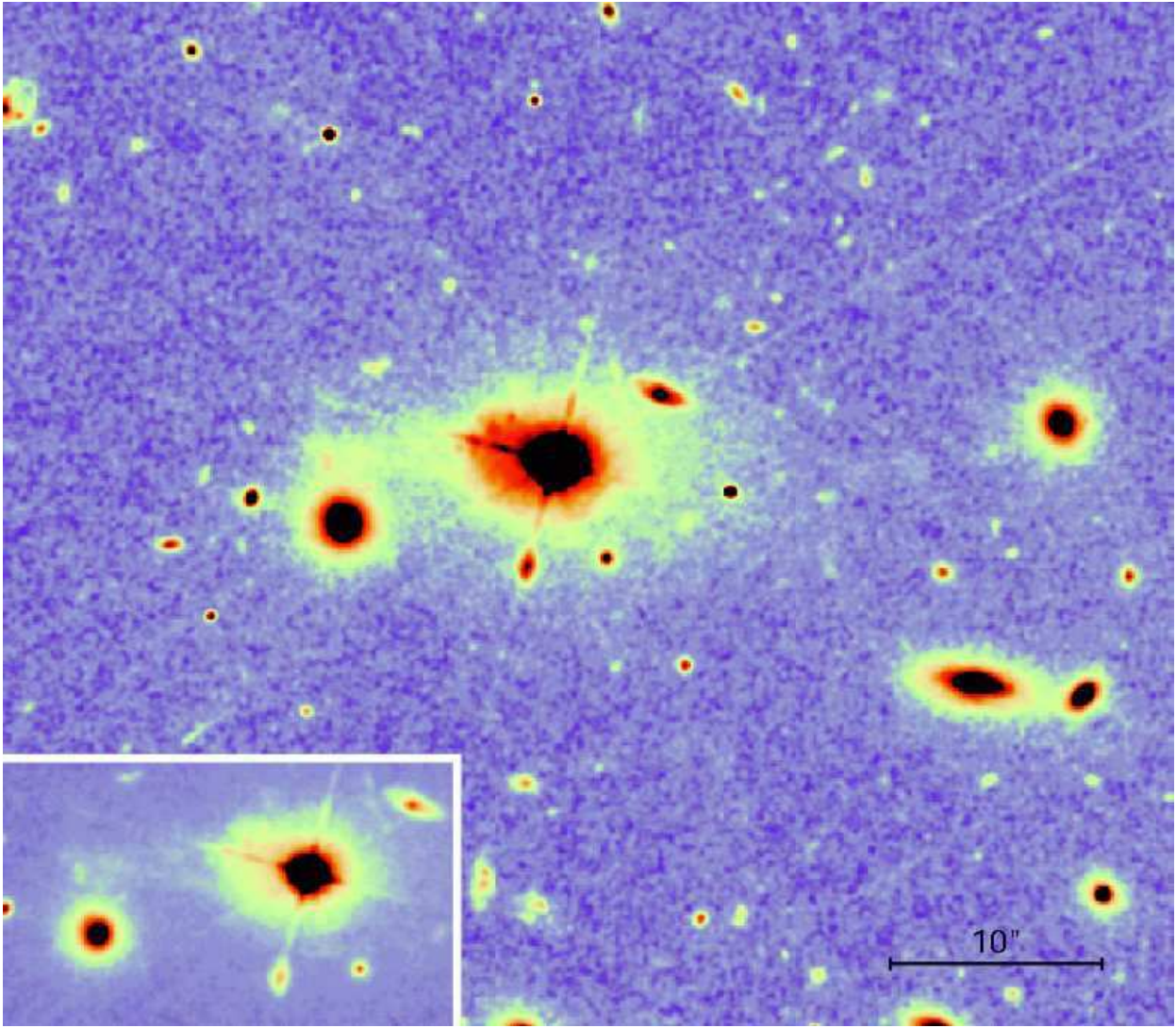}
\figcaption{{\it HST} WFC image of 4C\,37.43 in the F814W filter.  While this
filter emphasizes continuum emission, it also has $\sim10$\% of the peak
response at the wavelength of \ha\ in 4C\,37.43.  The inset shows a
lower-contrast version.  Note the filamentry structure in the ``bridge''
to the east of the quasar.\label{wfc_img}}
\end{figure}

\begin{figure}[!bt]
\epsscale{1.00}
\plotone{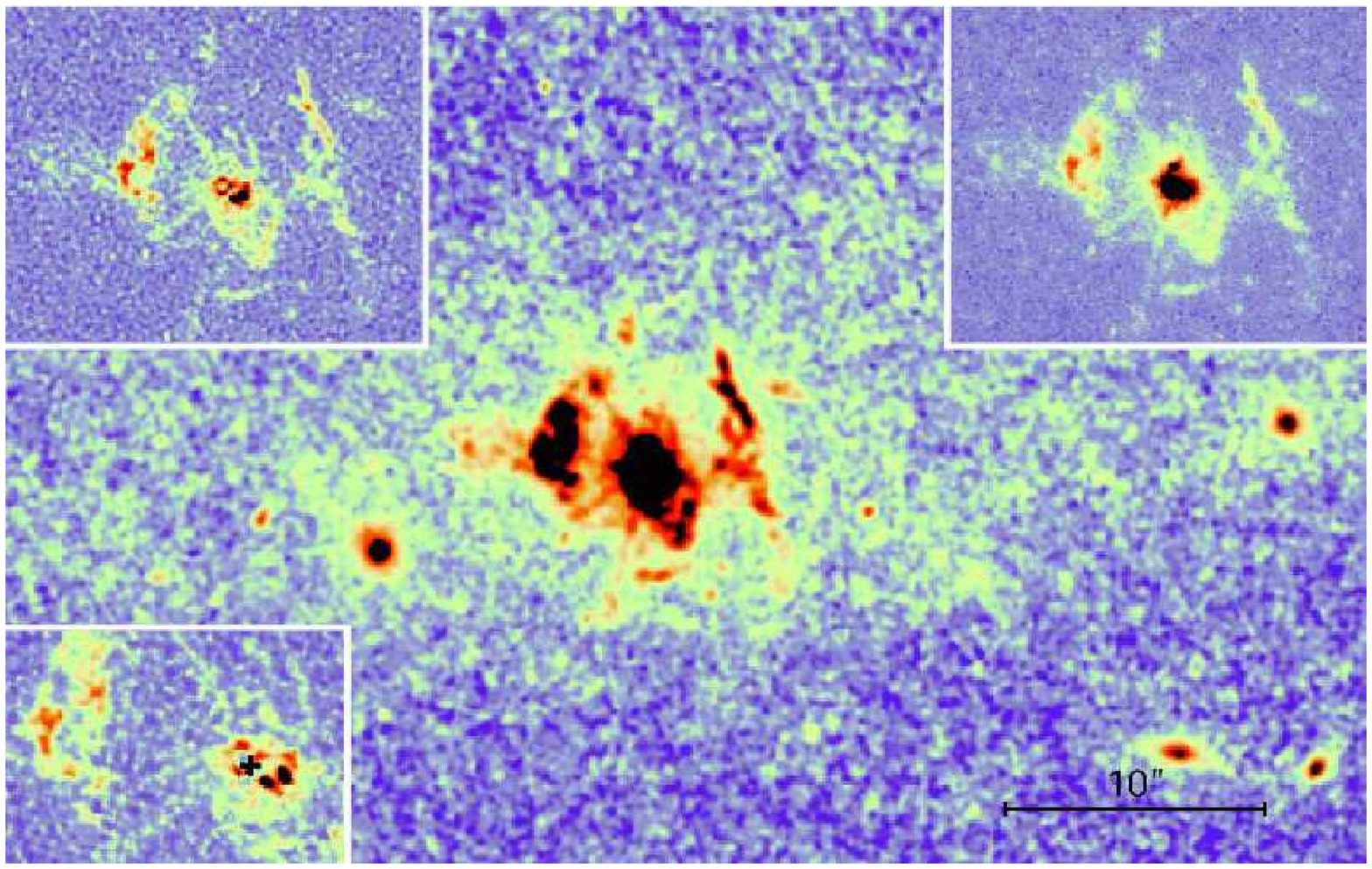}
\figcaption{{\it HST} WFC linear-ramp-filter (LRF) image of 4C\,37.43, with the
passband at the center of the image centered on \oiiilam. The upper-right
inset shows a lower-contrast version, and the upper-left inset shows
a {\it plucy} deconvolution and PSF removal.  Note the apparent bright
peak north of the quasar in the upper-right insert, which disappears in
the upper-left insert:  this is an LRF artifact \citep{can00a}.  The
lower-left insert shows a portion of the {\it plucy} deconvolution at
twice the scale of the rest of the figure.  Note that the two peaks of the
emission complex E1, $\sim4\arcsec$ east of the quasar, break up into a
large number of nearly unresolved knots.  The two bright
[\ion{O}{3}] emitting clouds just southwest of the quasar correspond to
the peak seen in our deconvolved ground-based image shown in
Fig.~\ref{gbimages}.
\label{oiii_img}}
\end{figure}

The {\it HST} images generally confirm and extend these results.  The WFC3 
%F814W continuum image is shown in Fig.~\ref{wfc_img}, the LRF \oiiilam\ image 
%in Fig.~\ref{oiii_img}, and the PC1 F814W image is included in Fig.~\ref{inner}.
F814W continuum image is shown in Fig.~\ref{wfc_img}, and the LRF \oiiilam\ 
image in Fig.~\ref{oiii_img}.  The relatively shallow PC F814W image shows
no significant new structure.

%\begin{figure}[!tb]
%\epsscale{.50}
%\plotone{f4.eps}
%\figcaption{{\it HST} [\ion{O}{3}] ({\it left}) and PC F814W ({\it right})
%images, showing the region immediately around the quasar after PSF
%removal.  The cross indicates the quasar position in the [\ion{O}{3}] image.
%The two bright emission knots to the west and southwest are weakly present
%in the PC ``continuum'' image, presumably because of \ha\ leakage through
%the filter.
%\label{inner}}
%\end{figure}
%
%
%\clearpage

The WFC3 F814W
continuum images are saturated at the center of the quasar, so it is difficult
to recover structure very close to the quasar by deconvolution or PSF fitting
and subtraction.  The PC1 F814W exposures were planned to avoid saturation
of the quasar; on the other hand, these images are strongly dominated by CCD
readout noise, so low-surface-brightness features have poor S/N.  The 
linear-ramp-filter \oiiilam\ images avoid saturation of the quasar and have
similar surface-brightness sensitivity to our best ground-based images.
However, the shift of the central wavelength with position on the CCD that
is a characteristic of these filters means that the \oiii\ emission more
than about 8\arcsec\ from the quasar will be lost.

The deep continuum image (Fig.~\ref{wfc_img}) shows that the bridge-like
structure to the east has a quite narrow core.  The high-surface-brightness
region northeast of the quasar, quite smooth at ground-based resolution, now
shows some knots that may be star-forming regions, although some of this
structure is due to regions of bright \ha\ emission leaking through the wings of
the F814W filter profile.

The \oiii\ image (Fig.~\ref{oiii_img}) shows an impressive amount of detail.
The two bright peaks comprising $E1$ break up into
a series of discrete knots, the long structure on the west side of the quasar
now has a braided appearance, and the very bright region just west of the
quasar is seen to have two discrete bright peaks; these last features are
confirmed in the PC image, where they are apparently due to the 
%\ha\ leak (Fig.~\ref{inner}).  The material to the southwest
\ha\ leak.  The material to the southwest
of the quasar forms an irregular ring, somewhat reminiscent of the one
seen in the extended emission region  around the quasar 4C\,25.40
\citep{sto87}.  However,
probably the most important new features are the faint linear structures
extending from the quasar to the northeast and southeast:  these are almost
certainly the edges of the ionization cone illuminating the bright emission
regions east of the QSO.  They are unlikely to be artifacts of the PSF, since
they persist undiminished after careful PSF subtraction, using a star
observed at the same detector position (Fig.~\ref{oiii_img}, upper-left
insert).  The presumed cone has an apparent opening angle of
$\sim92\arcdeg$ and its axis has a position angle of 77\arcdeg, not
particularly well aligned with the axis of the extended symmetric double
radio source, which has a position angle of $\sim108\arcdeg$ 
(\citealt{mil93}; see Fig.~\ref{gbimages}).  In retrospect, these linear 
features are confirmed in our deconvolved ground-based \oiii\ image in
Fig.~\ref{gbimages}, although the resolution there is certainly not
high enough to allow them to have been recognized as an ionization cone.

As mentioned above, at distances greater than a few arcsec, the LRF bandpass
shifts off the \oiiilam\ line, so some of the \oiii\ emission seen at
large distances east and west of the quasar in the our ground-based
\oiii\ image is not present in the {\it HST} image.  It is perhaps
significant that the axis defined by the most distant material seen in the
ground-based \oiii\ image (Fig.~\ref{gbimages}, bottom panels) lies close
to the axis of the ionization cone.  In fact, most of the extended emission
lies within the projected edges of this cone or those of an inferred counter 
cone, although some strong emission does not.  It is possible that this
emission lying outside the projected cone is evidence that the obscuring 
torus is leaky or irregular along some lines of sight.

\subsection{Reddening Along the Line-of-Sight}\label{reddening}

\begin{figure}[!tb]
\epsscale{0.5}
\plotone{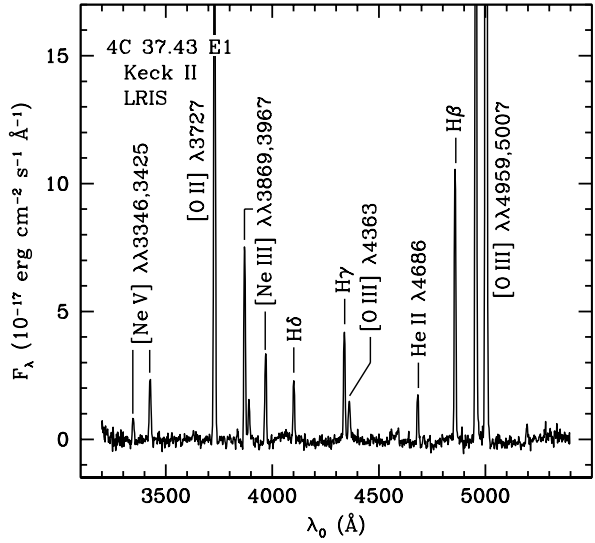}
\figcaption{Spectrum of 4C\,37.43 E1 obtained with Keck II and LRIS.
\label{lrisspec}}
\end{figure}

Before we can discuss emission-line ratios, we must estimate the reddening
along the line-of-sight from both Galactic dust and that associated with 
4C\,37.43 itself.  For the Galactic extinction, the dust column densities
found by \citet{sch98} imply 
$A_V=0.072$ mag.  After correcting the line fluxes by a standard Galactic 
extinction curve normalized to this value, we obtain an \ha\ to \hb\ flux
ratio of $3.58 \pm 0.29$ and an \hg\ to \hb\ flux ratio of $0.42 \pm 0.02$
(the former ratio was obtained from our spectrum from the UH 2.2 m telescope,
the latter from our Keck LRIS spectrum (Fig.~\ref{lrisspec}), 
which has a much better signal-to-noise
ratio).  Assuming that the reddening of the emission lines by intervening 
dust associated with 4C\,37.43 can be approximated by a standard Galactic 
reddening law \citep[\eg][]{car89}, we obtain intrinsic ratios 
$I($\ha$)/I($\hb$)=2.90$ and $I($\ha$)/I($\hb$)=0.47$ with $A_V=0.655$.  
These ratios
are quite consistent with Case B recombination, with little or no enhancement
of \ha\ due to collisional excitation.

\subsection{The Ly-$\alpha$/H$\alpha$ Ratio}\label{lyalpha}

\begin{figure}[!tb]
\epsscale{0.5}
\plotone{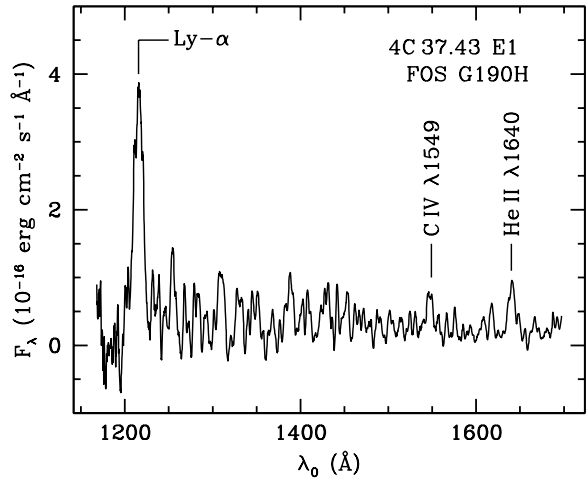}
\figcaption{Spectrum of 4C\,37.43 E1 obtained with {\it HST} and the
Faint Object Spectrograph.
\label{fosspec}}
\end{figure}

The UV spectrum of $E1$ obtained with the {\it HST} FOS is shown in
Fig.~\ref{fosspec}.  The observed flux in the \lya\ line is 
$5.3\pm0.6\times10^{-15}$ erg cm$^{-2}$ s$^{-1}$.
From our ground-based \ha\ imaging, using a
synthetic aperture designed to match as closely as possible the effective
FOS aperture, we obtain an \ha\ flux of 
$9.2\pm1.8\times10^{-15}$ erg cm$^{-2}$ s$^{-1}$, where most
of the quoted error is due to estimated uncertainties in the aperture
matching.  These values give a measured value of the \lya/\ha\ flux
ratio of $0.58\pm0.13$.  We must now attempt to correct for reddening.

Extrapolating the reddening found from the Balmer lines into the UV is
very uncertain.  If we simply assume that a standard Galactic reddening
law, with $R=3.1$, holds over the whole range, then the \lya\ to \ha\
flux ratio is $3.5\pm0.8$; if we consider values of $R$ ranging from
2.5 to 5 (roughly corresponding to the extreme values seen along lines of
sight in the Galaxy), we obtain flux ratios ranging from $7.5\pm1.7$ to 
$1.0\pm0.2$,
respectively.  The expected ratio under Case B conditions with no dust
in the emission region, $T_e=15000$ K, $N_e=300$, and uncomplicated geometries 
would be $\sim9$ (our more detailed photoionization models given in 
\S\ref{photoion} give a ratio of $\sim10$).  Thus, the intrinsic ratio 
could be consistent with little or no dust in the emitting region, or it 
could be a factor of $\sim10$ lower than this value, depending on the UV 
extinction properties of the {\it intervening}
dust.  Although it is often assumed that \lya\ emission is easily destroyed
by small amounts of internal dust because of the increase of effective
path before escape due to multiple scatterings, realistic simulations
indicate that rather large concentrations of dust are required to produce
significant deviations from the Case B ratio \citep{bin93}.  This result 
is largely due to the small gas column density in typical clouds, thus 
requiring few average scatterings before escape from the cloud. 
There can also be rather large variations in the \lya/\ha\ ratio because
of geometrical effects \citep{bin93}.  Our bottom line is that, if
something close to a standard Galactic reddening law with $R=3.1$
applies along the line of sight, both in our Galaxy and in the host
galaxy of 4C\,37.43, and if the individual emitting clouds are fairly
isotropic in their emission of \lya, then we infer substantial quantities
of internal dust.  However, the uncertainties in both the intervening 
reddening and the nature of the emitting clouds is such that we cannot 
exclude the possibility of little or no dust.

\subsection{Physical Properties of the Emission-Line Region}
\subsubsection{Average Electron Density and Temperature in the Emitting
Regions}

The luminosity-weighted average electron density in the [\ion{O}{2}]-emitting 
region can be determined
from the ratio of the \oiilam\ lines, and the electron temperature
in the [\ion{O}{3}]-emitting region similarly can be determined from the
ratio of the [\ion{O}{3}] $\lambda\lambda4363$,5007 lines \citep[\eg][]{ost89}.

\begin{figure}[!tb]
\epsscale{0.5}
\plotone{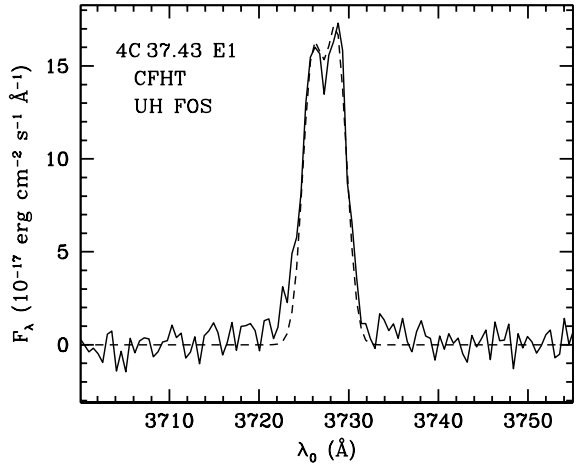}
\figcaption{The profile of the \oiilam\ doublet obtained with CFHT and the UH
Faint-Object Spectrograph.  The dashed line shows the best fit of
Gaussian profiles constrained to be centered on the expected wavelengths
of the doublet.
\label{oii}}
\end{figure}

Figure~\ref{oii} shows our high-resolution profile of the \oiilam\ doublet.
While the two lines are not quite resolved, the profile allows a
unique decomposition.  The fit shown is based on Gaussian profiles
placed at the laboratory wavelengths of the lines, fitted to the
deredshifted spectrum, where the redshift has been determined from
the [\ion{Ne}{3}] $\lambda3869$ line.  The ratio 
$I_{3729}/I_{3726}=1.03\pm0.03$, implying $N_e=380\pm60$ cm$^{-3}$ if
$T_e=10^4$ K, or $N_e=460\pm75$ cm$^{-3}$ if $T_e=1.5\times10^4$ K.
Earlier, \citet{ber87} had found $N_e=270$ by estimating the ionization
parameter from the emission-line spectrum and the incident flux from
the quasar luminosity and the projected distance to the emitting region.

The [\ion{O}{3}]-line ratio $I_{4363}/I_{5007}$ is well determined from
our Keck LRIS spectrum (Fig.~\ref{lrisspec}).  The observed ratio is
$0.0155\pm0.001$.  After correction for reddening, as determined according
to \S\ref{reddening}, we obtain $0.0178\pm0.001$ (we assume $R=3.1$ both
for the correction for Galactic reddening and for reddening at the
redshift of 4C\,37.43, but the difference over this wavelength range
from using any other reasonable value is similar to the random uncertainty).
For any likely value of $N_e$, $T_e=14500\pm400$ K.

\subsubsection{Comparison with Photoionization Models}\label{photoion}

It is clear even without running the models that no
single-parameter model can reproduce the observed spectrum: an ionization
parameter consistent with the [\ion{Ne}{5}]/[\ion{Ne}{3}] and 
[\ion{O}{3}] $\lambda5007$/H$\beta$ intensity ratios produces relatively
much weaker [\ion{O}{2}] $\lambda3727$ emission than is observed.
These and other inconsistencies between observed spectra of AGN narrow-line
and extended emission regions and computed photoionization
models based on a single ionization parameter
have led to several attempts to achieve better agreement by
considering at least two kinds of clouds with significantly different
physical properties.  One of the most successful of these approaches,
pioneered by \citet{vie92} and explored in considerable detail by 
\citet*{bin96},
involves using a combination of high-ionization-parameter matter-bounded
clouds and low-ionization-parameter ionization-bounded clouds.  The usefulness
of this approach was confirmed by \citet{rob00}, who found that such a
mixed-medium model not only gave a better fit to the emission-line spectrum
of extended emission in the radio galaxy 3C\,321 but also resulted in a more
reasonable value for the photoionizing flux of the (hidden) UV source.
Achieving good agreement with the observed emission spectrum
may also require the use of non-solar abundances \citep[\eg][]{vie92}.

We have run grids of models using the MAPPINGS3 photoionization code
\citep{dop95}.  Guided by the investigations just mentioned,
we have sought an acceptable model by (1) using two components, (2)
allowing either or both components to be density bounded, and (3)
allowing non-solar metallicities.  We estimate the flux density 
at the Lyman limit by making a plausible interpolation between the UV and 
X-ray luminosities given by \citet{lao94}, converting the luminosity density 
to our assumed flat cosmology with
$H_0=75$ km s$^{-1}$ Mpc$^{-1}$ and $\Omega_m=0.3$, and taking the actual
distance to $E1$ to equal the projected distance of 4\arcsec\ ($\sim19$ kpc).
We obtain the flux density from the quasar continuum at the Lyman limit
incident on $E1$ to be $I_{0}=3.3\times10^{-17}$ erg cm$^{-2}$ s$^{-1}$ Hz$^{-1}$.
We assume a two-component power-law photoionizing continuum, where
a component with index $\alpha=-1.5$ dominates in the UV and a flatter
component with index $\alpha=-1.2$ dominates in the X-ray region 
($f_{\nu}\propto\nu^{\alpha}$).  This continuum is normalized at the
Lyman limit to the value of $I_{0}$ found above.  We have also considered
a normalization to a flux density reduced by a factor of 2, corresponding
to a projection factor of $\sqrt2$ in the distance.  This reduced
incident flux results in best-fit models only slightly different from
those for unity projection factor (\ie\ using our assumed $I_{0}$), 
requiring minor changes to the densities and metallicity.

To obtain an acceptable fit to the observations, we find that
we need to have a metallicity significantly below solar, one component with
a density of several hundred cm$^{-3}$, and the other with a density near
1 cm$^{-3}$.  Neither component can be extremely optically thin at the
Lyman limit; on the other hand, we get slightly better agreement if the
low-density component is
somewhat density limited, with optical depths at the Lyman limit of
10 or so.  Our best agreement is for a model with about 1/3 solar metallicity,
with $\sim1/3$ of the flux of H$\beta$ coming from an essentially 
ionization-bounded 
component with average density $\sim500$ cm$^{-3}$, and $\sim2/3$ from a 
density-bounded component with average density $\sim2$ cm$^{-3}$ and an 
optical depth at the Lyman limit of 16.  The run of temperature in these 
regions is determined in a self-consistent manner from the models and averages
close to $10^4$ K in the high-density component and $1.5\times10^4$ K in the 
low-density component.  The average dimensionless ionization parameters are 
$U\sim6\times10^{-2}$ for the low-density region and $U\sim2\times10^{-4}$ for
the high-density region.  The results from two models, combining these two 
components in different ratios, are compared with the observed line ratios
in Table \ref{lineratio}.  Models 1 and 2 have, respectively, 25\% and 33\%
of the H$\beta$ flux coming from the high-density component.  Model 1 fits
most of the line ratios very well, but underestimates the [\ion{O}{2}] emission
by 40\%.  Model 2 fits all of the line ratios within $\sim20$\%.

\begin{center}
\begin{deluxetable}{lcccccccc}
\tablecaption{Observed and Modeled Line-Flux Ratios for 4C\,37.43 E1
\label{lineratio}}
\tablehead{\colhead{Line-Flux Ratios} & \colhead{[\ion{Ne}{5}]} & \colhead{[\ion{O}{2}]}
& \colhead{[\ion{O}{2}]} & \colhead{[\ion{Ne}{3}]} & \colhead{[\ion{O}{3}]} &
\colhead{\ion{He}{2}} & \colhead{[\ion{O}{3}]} & \colhead {[\ion{N}{1}]}\\
\colhead{} & \colhead{$\lambda3426$} & \colhead{$\lambda3726$}
& \colhead{$\lambda3729$} & \colhead{$\lambda3869$} & \colhead{$\lambda4363$} &
\colhead{$\lambda4686$} & \colhead{$\lambda5007$} & \colhead{$\lambda5199$}}
\startdata
Observed\tablenotemark{a} & 0.34 & 1.42 & 1.46 & 0.96 & 0.19 & 0.20 & 10.55 & 0.07\\
Model 1 & 0.34 & 0.87 & 0.90 & 0.85 & 0.19 & 0.20 & 10.31 & 0.07\\
Model 2 & 0.30 & 1.15 & 1.18 & 0.81 & 0.17 & 0.20 & \phn9.35 & 0.07\\
\enddata
\tablenotetext{a}{All line fluxes are given as ratios to the H$\beta$ flux}
\end{deluxetable}
\end{center}

From our estimate of the dereddened \ha\ flux (\S\ref{lyalpha}), an estimate
of the volume over which this flux is emitted, and the relative amounts of
Balmer emission from the two components of our photoionization model,
we can estimate the filling factors of the two density regimes.
If $f_{\rm H\beta}$ is the observed H$\beta$ flux from a given region,
the filling factor $\phi$ can be written as
$$
\phi=\frac{3f(1+z)^6}{4\pi j_{\rm H\beta} \theta^3 d_L}{\rm ,}
$$
where $j_{\rm H\beta}$ is the emission coefficient for H$\beta$,
$\theta$ is the angular radius of a sphere having the same volume as the 
emitting region, and $d_L$ is the luminosity distance.  The quantity
$j_{\rm H\beta}/N_pN_e$ is nearly independent of density over the
relevant density range and can be
taken to be a function of temperature only \citep[\eg][]{ost89}.
For the case of $E1$, if we assume that $\theta\approx0\farcs75$ and
that $N_p\approx N_e$, with values of 2 cm$^{-3}$ and 500 cm$^{-3}$ for the 
low-density and high-density regions, respectively, we then obtain
filling factors $\phi_2\sim0.4$ and $\phi_{500}\sim2\times10^{-5}$.  
Given the
strong sensitivity of these calculations to the size of the emitting region,
along with the simplicity of our model, these results likely point to
a situation in which the low-density, \oiii-emitting gas
fills the volume, in which are embedded tiny, dense regions of \oii-emitting
gas.  Normalizing to the \ha\ flux over a $1\farcs4\times4\farcs2$ region
centered on $E1$, we obtain $\sim4 M_{\odot}$ of gas in the low-density
component and $\sim0.05 M_{\odot}$ of gas in the high-density component.

While our two-component model is clearly a great oversimplification, it can
still give some qualitative insight into the properties of the emission-line 
region.  It is clear that most ($\sim90$\%) of the \oiii\ emission comes from
regions of low density, while the \oii\ emission comes almost entirely from
regions with densities $\sim200$ times higher.  This means that attempts
to estimate the ionization parameter (and thence the pressure) from
the [\ion{O}{2}]/[\ion{O}{3}] ratio \citep[\eg][]{fab87,cra00} are not likely to
be reliable.  As it happens, we do indeed find a pressure of 
$5\times10^6$ cm$^{-3}$ K for the high-density regions, close to the average 
pressure found by \citet{cra00} for the EELR around 4C\,37.43.  But the 
pressure in the low-density region, which is arguably the component that is
likely to be in pressure equilibrium with a hot external medium, is 
$\sim3\times10^4$ cm$^{-3}$ K, over 2 orders of magnitude lower.

The high-density regions are thus clearly not in pressure equilibrium
with their surroundings. Because they are located
some 20 kpc (in projection) from the quasar and do not show a strong
correlation with the continuum structure of the host galaxy (and are hence
unlikely to be gravitationally confined), they will dissipate on a timescale
on the order of the sound-crossing time.  Our photoionization calculations
indicate that these dense clouds (or at least the ionized part of them)
are typically only about 0.1 pc thick, so, with a typical sound speed of
$\sim15$ km s$^{-1}$, they are likely to have lifetimes of $\lesssim10^4$
years.  It seems most likely, then, that these clouds are continuously
regenerated by shocks propagating through the surrounding medium.  
On the other hand, the relative weakness of UV lines such as \ion{O}{4}
$\lambda1402$ and \ion{C}{4} $\lambda1549$ with respect to \ion{He}{2}
$\lambda1640$ indicates that photoionization by the central QSO strongly
dominates over ionization by the shocks themselves in generating the 
observed emission lines.

\subsection{Kinematics of the Ionized Gas}
\begin{figure}[p]
\epsscale{1.0}
\plotone{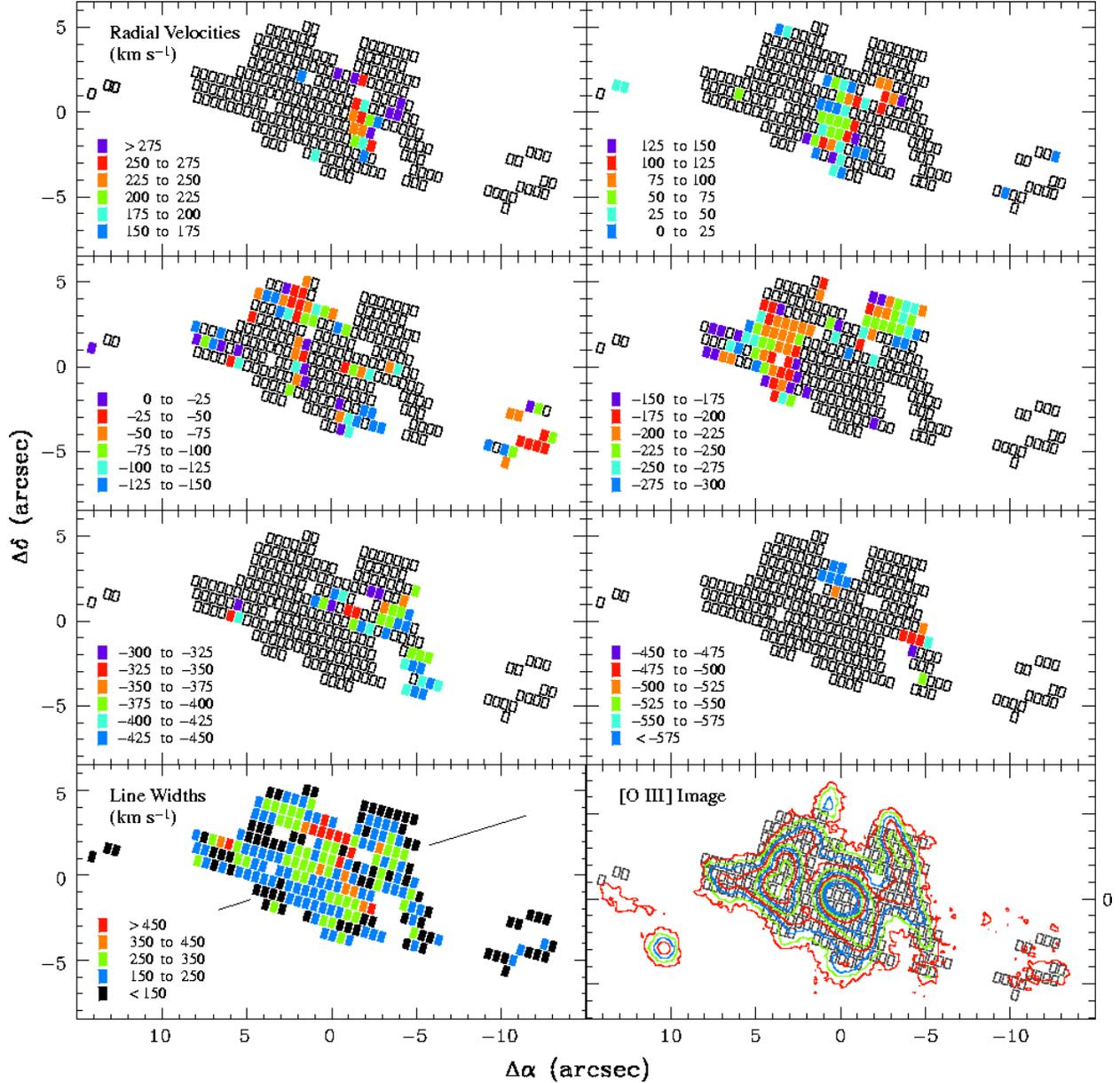}
\figcaption{The upper six panels give radial velocities, in km s$^{-1}$, 
relative to that of the nuclear narrow-line
region, for \oiii\ emission at different points in the EELR around 4C\,37.43.
Each of the 6 panels in the top part of the figure shows a different 150
km s$^{-1}$ range, and the different colors indicate emission in different
25 km s$^{-1}$ intervals within each of these ranges.  Thus the gross 
characteristics of the velocity field can be judged by comparing these
six maps, and finer details can be seen within each individual map. Note 
that at some points two or three velocities may be present along the same 
line of sight.  The lower left panel shows velocity widths of the \oiii\
line, again given in km s$^{-1}$.  This panel also indicates the radio axis.
The panel at the bottom right shows the location of the radial
velocity and line-width measurements superposed on a contour representation of
our groundbased \oiii\ image.
\label{oiiivel}}
\end{figure}

Our moderate-resolution image-sliced spectroscopy of the \oiii---H$\beta$
region covers most of the obvious emission seen in our deep narrow-band 
\oiii\ imaging.  Velocities (relative to the nuclear narrow-line region) 
measured from the \oiiilam\ line are shown in Fig.~\ref{oiiivel}.

Our velocity map is in general agreement with those of \citet{dur94} and
\citet{cra00}.
The relative velocities range from $-700$ to $+325$ km s$^{-1}$.  One's
general impression is that the velocity field is locally ordered but
globally disordered:  individual filaments and components show continuous
velocity gradients, but the velocities from region to region do not
suggest any sort of simply organized kinematics.  Note that the brightest
regions on both the east and west sides of the QSO have substantial negative
velocities.

\citet[][p.~442]{cra00} mention that ``...it is notable that the region of largest
blueshift and of higher ionization in both the eastern and western clouds,
and the highest FWHM in the western cloud (but not the eastern cloud) are
all where the optical gas crosses the radio source axis.'' Raising the
possibility of a jet-cloud interaction, they however point out that
``...there is no obvious distortion in the radio source to support this
interpretation.''  As Fig.~\ref{oiiivel} shows, while some fairly high-velocity
gas lies close to the radio axis on the western side, gas at similar velocities
exists at many locations, and the highest velocities are found elsewhere.
In any case, it would seem difficult to understand why an interaction
between the radio jet and the gas should give large negative radial velocities
for {\it both} the approaching and the receding jet.  We cannot make detailed 
comments on the variations in ionization because in many areas the \hb\
emission is too weak to measure.  What measurements we do have show that
the peak \oiii/\hb\ ratio  falls near the peaks of the \oiii\ intensity,
significantly to the north of the radio jet on both the east and west sides.
In fact, our measurement of the \oiii/\hb\ ratio at the brightest region
crossed by the jet on the east side is 5.3, the lowest we are able to 
measure anywhere, compared with 11.0 about 3\farcs5 north of this position.

We show our measurements of the \oiii\ line width in the lower left map in
Fig.~\ref{oiiivel}.  These measurements must be taken with some caution,
since multiple velocities are present along many lines of sight, and it
may not always have been possible to deconvolve indivdual components.  With
this proviso, there seems to be some enhancement in the line width near
where the radio jet crosses the linear structure on the west side, but
similar line widths are seen at many other locations in the nebula.  The
largest line widths seem to be associated with the high velocity gas about
2\arcsec\ north of the quasar.

Overall, then, we find no compelling evidence for a significant effect on
the gas kinematics or ionization from an interaction with the radio jet.

\section{Origin of the Extended Gas}

\subsection{Varieties of QSO EELRs}
Considering QSO EELRs as a class, we can distinguish at least four 
different types.  The first are those in which the emission-line gas
is closely associated with the morphology of the host galaxy; apparently,
cold disk gas in the host galaxy is photoionized by the QSO nucleus.
Such examples among QSOs (as contrasted with Seyferts) are rare, possibly
because few QSOs have host galaxies with intact cold disks, and the
ionization cones are likely not to intercept the disks in any case.
The only likely example we are aware of is PG\,0052+251 \citep{sto87}.

The second type is that in which the emission-line gas appears to be
connected with the radio morphology.  At least among
low-redshift QSOs, strong emission associated with radio structure is
not common, but we do have an example in PKS\,2251+11, where the two
brightest peaks in the extended emission bracket the southeast radio lobe
\citep{sto87,dur94,cra00}.  In such cases, it is likely that the emitting 
gas is either
ambient material that has been shocked by the radio jet, or that it is
gas from the host galaxy that has become entrained in the radio jet.

A third type includes cases for which a neighboring gas cloud, confined
either gravitationally (as in a companion galaxy) or by thermal pressure
(as in a cooling flow, as envisioned by \citealt{fab87}), lies within the
ionization cone of the QSO and is close enough to be photoionized by the
QSO's UV continuum.  One trivial example of this type is a small galaxy
$\sim23$\arcsec\ west of Mrk\,1014. \citet{sto87} noticed that this galaxy was
enhanced in the narrow-band image centered on the wavelength of \oiiilam\
in Mrk\,1014, and recent spectroscopy confirms strong [\ion{O}{3}] emission
from this object at the redshift of Mrk\,1014 \citep{can00b}.  An 
intriguing possibility is that ionization by nearby QSO may make visible
objects that would be very difficult to detect by any other means.
There is reasonable evidence that dwarf galaxies are heavily dark-matter
dominated (\eg\ \citealt{mat98}, and references therein), so such galaxies can, 
in principle, trap gas in potential wells that are not well advertised 
in terms of their stellar
luminosity.  The ubiquitous dwarf star-forming galaxies ({\it aka}
extragalactic \ion{H}{2} regions) are a case in point:  they cannot have
sustained their present rate of star formation for more than a small
fraction of a Hubble time.  Many of these must have been virtually
invisible in their earlier quiescent state; yet they have to have had
substantial reservoirs of gas at moderately high densities in order to
allow the current star formation to proceed.
Objects such as these, with few stars but some gas in a largely dark
matter potential well, would normally be undetectable at moderate redshifts,
but they would become quite noticeable if the gas were to be ionized by
a nearby QSO.  This might be one way of detecting at least one class
of the ``ghost galaxies'' proposed by Kormendy \& Freeman (1998;
see also \citealt{tre01}).
A possible example of such a case
is the discrete [\ion{O}{3}]-emitting object near NAB\,0205+05
\citep{sto87}, for which the equivalent width of the \oiiilam\ line is
extremely high.

Finally, there is the type of EELR which is closely associated
with the QSO, but for which the ionized gas, though often having a highly
structured morphology, shows no obvious relation to either the radio or
the optical continuum morphology.  Most QSO EELRs seem to be of this type,
although few have the luminosity or show the richness of structure of
that of 4C\,37.43.

\subsection{Observational Constraints from the 4C\,37.43 EELR}
To recapitulate our main results from \S\ref{photoion}, the region of the 
4C\,37.43 EELR that we have 
investigated appears to comprise two main regimes, one with a density of
$\sim2$ cm$^{-3}$ and a temperature of $\sim1.5\times10^4$ K, the other 
with a density of $\sim500$ cm$^{-3}$ and a temperature of $\sim10^4$ K.  The
filling factor of the low-density gas is close to unity, while that of the
higher-density gas is $\sim10^{-5}$.  It is possible that the low-density
gas is close to pressure equilibrium with a surrounding hot external medium, but
the high-density gas cannot be.  The relatively low pressure of 
$\sim3\times10^4$ cm$^{-3}$ K that we deduce for the low-density gas 
argues against a cooling flow origin for the emission-line region of 4C\,37.43,
as proposed by \citet{cra00}.  

The referee has pointed out that photoionization of a typical giant 
molecular cloud (GMC) complex by a quasar could produce a similar
two-phase density structure, as the photoionized region eats into the dense
cloud and the ionized gas rapidly expands to reach pressure equilibrium with
the surrounding medium.  This is an attractive suggestion, since GMCs
are self-gravitating and could be stable over  timescales of $>10^8$ years.
However, as we have noted above, there is virtually no correlation
between host galaxy morphology and the distribution of the gas in the EELR
for 4C\,37.43 and similar objects; in these cases, the quasar cannot simply
be ``lighting up'' {\it in situ} interstellar clouds.  The brightest emission 
regions around 4C\,37.43 are mostly at projected distances of $\sim20$ kpc,
where one would not normally expect to find dense molecular clouds.
Furthermore, strong extended emission is not the norm in QSOs,
even amongst the steep-radio-spectrum classical double sources for which
it is statistically more likely.  Why do a few objects like 4C\,37.43 have
such prolific displays of extended emission, while the majority of apparently
similar objects show little or none? As we have previously noted \citep{sto87},
there is an apparent correspondence between
the incidence of strong extended emission among QSOs and the presence of
overt signs of strong interaction, such as close companion galaxies and
tails or bridges seen in continuum emission.  This is certainly true in
the case of 4C\,37.43, where the continuum asymmetry and bridge- or tail-like
structure to the east strongly suggest a recent merger.  
One might therefore suppose
that GMCs could have been tidally ejected to large distances and out of
the plane of the original disk in which they were formed.  But GMCs are
sufficiently dense that their trajectories should not be significantly
affected by hydrodynamic interaction with the ambient gas, and they should
follow essentially the same paths the stars do.  Once again, the general 
lack of correspondence between the continuum and emission-line morpholgies in
4C\,37.43 argues against this scenario.  

Because the high-density clouds cannot be confined by external pressure,
and gravitational confinement seems unlikely, they have probable lifetimes of 
$\lesssim10^4$ years. It therefore seems likely that they 
must be continuously regenerated by shocks.  
If so, it remains a possibility that the shocks
responsible for the dense regions are simply due to collisions of sheets of
gas during the merger \citep{sto90}.  The
velocities of up to 700 km s$^{-1}$ that we see are a significant, but perhaps 
not completely fatal, objection to this picture.  It is quite possible that
much of the extended ionized gas we see was originally ejected from one or
both of the galaxies by tidal forces during the merger.
However, it now seems likely to us
that the origin of the shocks needed to produce the high-density clouds is an 
\textit{indirect} result of the merger: a starburst-driven galactic
superwind.

Virtually all ultra-luminous infrared galaxies (ULIGs) are starbursts
triggered by major mergers \citep[\eg][]{san96}.  At least some significant 
fraction of these also have QSO nuclei \citep[\eg][]{lut99}, which must also 
have been triggered by the merger \citep{can01}.  The ULIG phase can dominate
the IR emission for up to $\sim300$ Myr \citep{can01}; if the QSO has a 
lifetime greater than this, some objects identified as ``ordinary'' QSOs
(\ie\ those with no evidence of excessive FIR emission) will
have previously been ULIGs as well.  One of the consequences of a vigorous
starburst is a galactic superwind \citep[and references therein]{hec01}, 
which will interact with the surrounding medium and particularly with
any tidally ejected gas.  This superwind provides a mechanism for producing 
shocked gas, potentially at large distances from the QSO\footnotemark[6].
\footnotetext[6]{This possibility was first mentioned to us by Mike Dopita.}
It also may be responsible for entraining any ambient cool gas 
(including that possibly ejected during the encounter) and controlling
both its location and its velocity.
This picture is consistent with both the distribution and the velocities
of the ionized gas we see in 4C\,37.43.  The main emission regions are found 
at roughly the same projected distances to the east and west of the quasar, 
and knots of emission are arranged along linear or arc-like structures 
roughly perpendicular to the direction to the quasar, which plausibly 
correspond to surfaces of the expanding bubbles.  As can be seen from
Fig.~\ref{oiiivel}, the bulk of the gas in the most luminous emission regions
has projected velocities between about $-200$ and $-300$ km s$^{-1}$. 
There are similar positive velocities in regions with weaker emission;
also in weaker-emitting regions, there are negative velocities ranging up
to $-700$ km s$^{-1}$.  The predominance of negative velocities could
simply be due to chance: the uneven distribution of the gas (particularly if it
is mostly tidal debris), coupled with the illumination pattern of the UV
continuum from the quasar.  However, if the velocities have a significant
outflow component, a bias towards negative velocities
could also be a natural consequence of screening by dust blown out by
a superwind.  Also, the tendancy for the highest velocities to be
found amongst the weaker-emitting components is consistent with the
expectation that initially less dense and less massive clouds would 
be accelerated to higher velocities.

Furthermore, of the dozen most luminous EELRs in the survey of low-redshift
QSOs by \citet{sto87} (including, of course, that of 4C\,37.43), two are
known to be associated with QSOs having ultraluminous IR host galaxies
having confirmed starburst or recent post-starburst stellar populations
(3C\,48, \citealt{can00a}; Mrk\,1014, \citealt{can00b}).  This association
lends support to a connection between the EELR and starburst activity.  
\citet{can01} have suggested that
QSOs that are also powerful radio sources, such as 3C\,48, can break
through a dust cocoon associated with the starburst more rapidly than
can most radio-quiet QSOs, allowing the central UV source to illuminate
and photoionize the extended gas early on.  If this is true, it could
explain why the most luminous EELRs are associated with powerful extended
radio sources \citep{bor82,bor84,sto87}.

\section{Summary}
The observations presented here have given us the clearest view yet of an
extended emission region around a quasar.  The \textit{HST} WFPC2 imaging
in the \oiii\ emission line shows the first clear evidence of an ionization
cone on the east side of the quasar and confirms the presence of very
high surface brightness emission $\sim0\farcs5$ west of the quasar.
Much of the emission is arrayed in linear or arc-like structures, many of which
appear to lie about 4\arcsec\ from the quasar and to be centered on it.
The spectrum of the bright emission complex $E1$ is consistent with
photoionization by the quasar, provided that the region comprises two
main components, both having metallicities around 1/3 solar:  
(1) a low-density ($\sim2$ cm$^{-3}$) gas with essentially
unity filling factor, and (2) a high-density ($\sim500$ cm$^{-3}$) gas with
a very low ($\sim10^{-5}$) filling factor and distributed in thin ($\sim0.1$
pc) filaments or sheets.  The very short ($\sim10^4$ year) lifetimes of
these dense regions suggest their origin in shocks.
From the spectroscopic and imaging evidence, together with our knowledge 
that at least some
QSOs have gone through a strong starburst phase as a result of major mergers,
we believe that it is quite likely that such shocks are consequences of
superwinds driven by a recent starburst in the quasar host galaxy.

\acknowledgments

We thank Gabriela Canalizo for assisting with the Keck LRIS observations,
Geoff Bicknell, Mike Dopita, and Josh Barnes for helpful discussions, and 
Mike Dopita,
Lisa Kewley, and Ralph Sutherland for help with running the MAPPINGS 
photoionization code.  We also thank the anonymous referee for a careful
reading of the paper and for comments that helped us improve the 
presentation. Support of the HST observations
was provided by NASA through Grant Nos. GO-3538.01-91A and GO-06490.01-95A
from the Space Telescope Science Institute, which is operated by AURA, Inc.,
under NASA Contract No.~NAS\,5-26555.  Some of the ground-based observations
were supported by NSF Grant No.~AST\,95-29078. Some of the data presented 
herein were obtained at the W.M. Keck Observatory, which is operated as a 
scientific partnership among the California Institute of Technology, the 
University of California and the National Aeronautics and Space Administration,
and which was made possible by the financial support of the W.M. Keck 
Foundation.  The authors recognize the very significant cultural role that 
the summit of Mauna Kea has within the indigenous Hawaiian community and
are grateful to have had the opportunity to conduct observations from it.

\end{document}